\documentclass[a4paper,11pt]{article}
\usepackage{pos}
\usepackage{tikz}

\title{Hunting light Higgses at the LHC in the context of the 2HDM Type-I}

\author[a,b]{S.~Moretti}
\author*[a,c]{S.~Semlali}
\author[c]{C.~H.~Shepherd-Themistocleous}

\affiliation[a]{School of Physics and Astronomy, University of Southampton, Southampton, SO17 1BJ, United Kingdom}
\affiliation[b]{Department of Physics \& Astronomy, Uppsala University, Box 516, SE-751 20 Uppsala, Sweden}

\affiliation[c]{Particle Physics Department, Rutherford Appleton Laboratory, Chilton, Didcot, Oxon OX11 0QX, United Kingdom}

\emailAdd{s.moretti@soton.ac.uk and stefano.moretti@physics.uu.se}
\emailAdd{souad.semlali@soton.ac.uk}
\emailAdd{claire.shepherd@stfc.ac.uk}

\abstract{We show the reinterpretation of existing searches for exotic decays of the Standard Model (SM)-like Higgs, $H \to aa (hh)$, in various final states, in the framework of the 2-Higgs Doublet Model (2HDM) Type-I.  We then explore a new search for such light Higgses, $a$ and $h$, at the Large Hadron Collider (LHC) Run 3 for an integrated luminosity of 300 $\text{fb}^{-1}$. After performing a scan over the model parameters, we found that the inverted scenario of Type-I offers a new promising signal in the form of the following cascade decays: $H \to Z^{*}a \to Z^{*}Z^{*}h \to b\overline{b} \mu^- \mu^+ jj$. We investigate then its significance through a full Monte Carlo (MC) simulation down to the detector level.}

\FullConference{%
  41st International Conference on High Energy physics - ICHEP2022\\
  6-13 July, 2022\\
  Bologna, Italy
}

\newcommand{\eslash}{\ensuremath{{\hbox{$E$\kern-0.6em\lower-.05ex\hbox{/}\kern0.10em}}}}
\begin{document}
\maketitle

\section{Introduction}
The current measurements of the observed Higgs couplings~\cite{CMS:2022dwd,ATLAS:2022vkf} at the LHC leave room for non-SM decays. The ATLAS and CMS collaborations have constrained these to be below 0.14~\cite{ATLAS:2022yvh} and 0.18~\cite{CMS:2022qva}  at 95\%, respectively. Among the prominent processes, there are searches for exotic Higgs decays to a pair of light scalars or pseudoscalars, $H\to hh$ and $aa$, respectively, which then decay to SM particles. These processes are well motivated by many theories, which are consistent with the LHC measurements and foresee the presence of extra light scalars/pseudoscalars, such as the 2HDM. Herein, we briefly review some of the reported searches for exotic Higgs decays in various final states in the framework of 2HDM Type-I. While doing so, we have come across a new promising signature to search for light Higgses in the cascade decay $H\to aZ^{*} \to h Z^{*}Z^{*}\to \mu^+\mu^-jjb\overline{b}$. We then investigate its feasibility at the LHC at a centre-of-mass energy of 13 TeV and an integrated luminosity of 300 $\text{fb}^{-1}$~\cite{Moretti:2022fot}.

\section{Review of 2HDM-I}
The scalar sector of $\mathcal{CP}$ conserving 2HDM consists of two $SU(2)_L$ doublets with hypercharge $Y=1$, which can be parametrised as follows:
\begin{align}
H_i = \left( \begin{array}{c} \phi_i^+ \\ \frac{1}{\sqrt{2}} (v_i +
\rho_i + i \eta_i) \end{array} \right)~~ \text{with} ~~\  \ (i=1, 2),
\end{align}
where $v_{1}$ and $v_2$ are the Vacuum Expectation Values (VEVs) of the two doublets $H_1$ and $H_2$.
The general $SU(2)_L\times U(1)_Y$ invariant scalar potential is given by:
\begin{eqnarray}
V(H_1,H_2) &=& m_{11}^2 (H_1^\dagger H_1)+m_{22}^2(H_2^\dagger H_2)-[m_{12}^2(H_1^\dagger H_2)+{\rm h.c.}] 
+ \frac{\lambda_1}{2}(H_1^\dagger H_1)^2
+\frac{\lambda_2}{2}(H_2^\dagger H_2)^2 \nonumber \\
&+&\lambda_3(H_1^\dagger H_1)(H_2^\dagger H_2)
+\lambda_4(H_1^\dagger H_2)(H_2^\dagger H_1)+ \left\{\frac{\lambda_5}{2}(H_1^\dagger H_2)^2+ \rm h.c. \right\}.
\label{eq2}
\end{eqnarray}
After the Electro-Weak Symmetry Breaking (EWSB), eight Higgs degrees of freedom ($\rho_i,~\eta_i~\text{and}~\phi_i^\pm$) manifest  as five physical Higgses: two $\mathcal{CP}$-$even$, $h$ and $H$ with $m_h < m_H$, one $\mathcal{CP}$-$odd$ ($a$) and a pair of charged Higgs $(H^\pm)$. The  three remaining degrees of freedom correspond to the Goldstone bosons, $G^\pm$ and $G^0$ are absorbed by the longitudinal components of vector bosons $W^\pm$ and $Z$, respectively. Extending the $Z_2$ symmetry to the Yukawa sector to avoid tree-level Higgs-mediated Flavour Changing Neutral Currents (FCNCs) leads to four various structures of Higgs-fermion interactions: Type-I, Type-II, Type-X (or Lepton specific) and Type-Y (or Flipped). Note that this symmetry is softly broken by the term $[m_{12}^2(H_1^\dagger H_2)+{\rm h.c.}]$ in Eq.~(\ref{eq2}). Here, we will only focus on Type-I where one of the doublet couples to all fermions, thus the Higgs-fermion interactions are flavour diagonal in the mass eigenstate basis. Tab.~\ref{tab1} illustrates the different couplings in the framework of Type-I in terms of the mixing angle $\alpha$ and $\beta$, with $\tan \beta = \frac{v_2}{v_1}$.
		\begin{table}[h!]
	\begin{center}
		\resizebox{0.38\textwidth}{!}{
			\begin{tabular}{|c |c|c|c|} \hline
				Higgs & $up$-type & $down$-type & leptons\\ 
				\hline \hline 
			    h & $c_\alpha/ s_\beta$ & $c_\alpha/ s_\beta$ & $c_\alpha/ s_\beta$ \\
			    \hline 
			    H & $s_\alpha/ s_\beta$ & $s_\alpha/ s_\beta$ & $s_\alpha/ s_\beta$ \\
			    \hline
			    A & $-\cot \beta$  & $\cot \beta$  & $\cot \beta$ \\
			    \hline
			\end{tabular} 		
		}          
	\end{center}
\caption{Higgs-fermions couplings in the 2HDM Type-I}
	\label{tab1}
\end{table}	 
\vspace{-0.2cm}
\subsection{Theoretical and Experimental Constraints}
Different numerical tools are used to consider theoretical requirements and experimental constraints in our scans. We use \texttt{2HDMC-1.8.0}~\cite{2hdmc} to scan over the parameter space of the 2HDM and to check that each point satisfies constraints from EW precision tests ~\cite{particle2020review}. \texttt{HiggsBounds-5.10.0}~\cite{Bechtle:2020pkv} and \texttt{HiggsSignals-2.6.2}~\cite{Bechtle:2020uwn} are called to enforce constraints arising from direct Higgs searches at colliders and from Higgs boson current signal strength measurements. Constraints from flavour physics observables are tested by \texttt{SuperIso}~\cite{Mahmoudi:2008tp}. 
Furthermore, additional constraints from recent searches for exotic Higgs decays $H \to aa \to \mu^+\mu^-b\overline{b}~\text{\cite{CMS:2018nsh,ATLAS:2018emt}}$, $\tau^+\tau^- \mu^+ \mu^-~\text{\cite{CMS:2018qvj}}$ and $b\overline{b}\tau^+\tau^-~\text{\cite{CMS:2018zvv}}$ in the mass range $[15,~62.5]$ GeV at $\sqrt{s}=13~$ TeV are considered. 

\section{Numerical Analysis}
In this analysis, we confine ourselves to the inverted hierarchy scenario, where the heaviest Higgs is identified as the  125 GeV state observed at the LHC. We perform a systematic scan over the following ranges:
\begin{eqnarray*}
	m_h \in [10~\text{GeV},~90~\text{GeV}],~~~m_H = 125~\text{GeV},~~~m_a \in [10~\text{GeV},~90~\text{GeV}], \nonumber \\
	m_{H^\pm}\in [100~\text{GeV},~160~\text{GeV}],~~~\tan\beta \in  [2.5,~25],~~~\sin(\beta-\alpha) \in [-0.7,~0.0],
	\label{eq1}
\end{eqnarray*}
with $m_{12}^2$= $m_a^2\tan\beta/(1+\tan^2\beta)$. 

	\begin{figure}[!h]
	\centering
	\resizebox{0.55\textwidth}{0.21\textheight}{
		\includegraphics{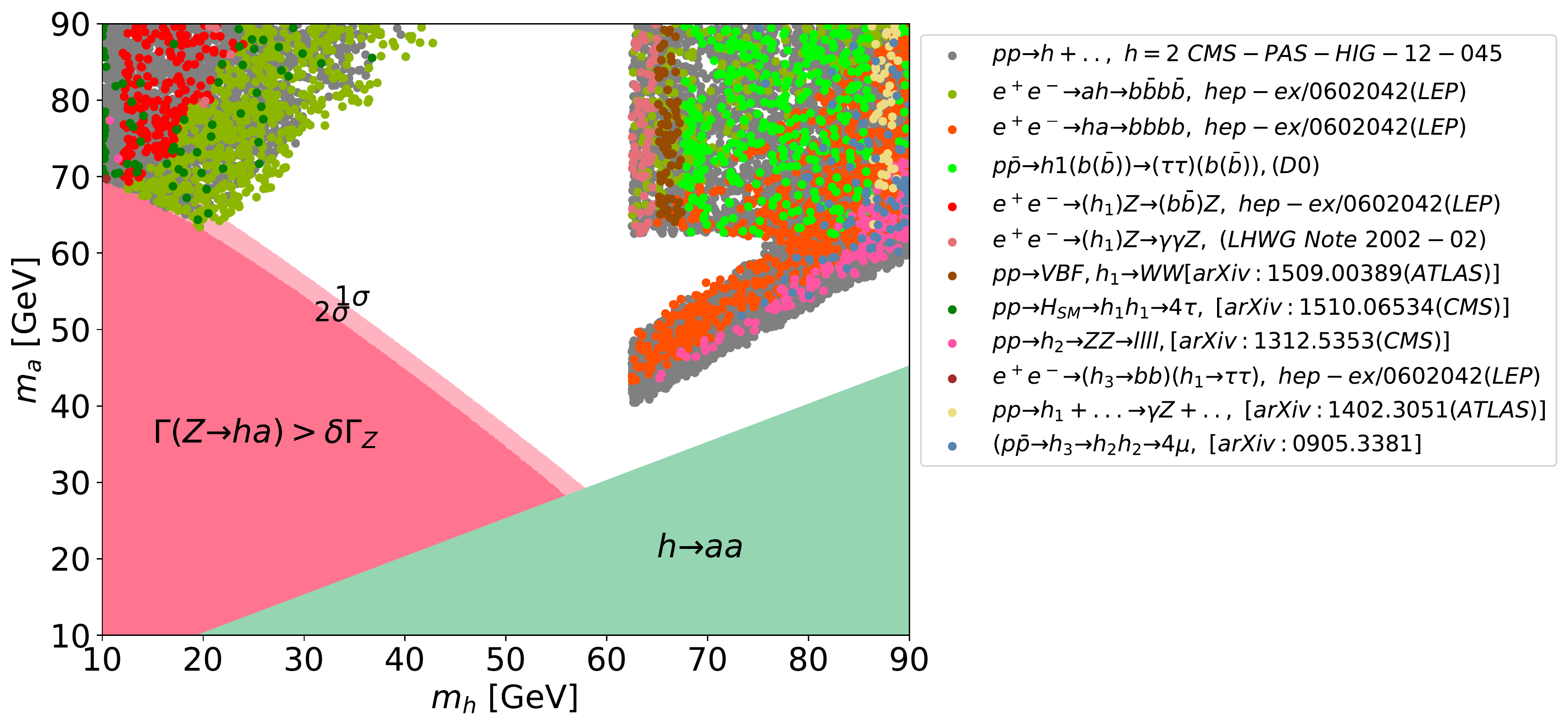}}
	\caption{Allowed parameter space in the 2HDM Type-I  at 95\% C.L. Coloured dots represent the searches to which the relevant ($m_h, m_a$) regions are sensitive to}
	\label{fig1}
\end{figure}

\begin{figure}[h!]
	\centering	
	\resizebox{0.42\textwidth}{0.23\textheight}{\includegraphics{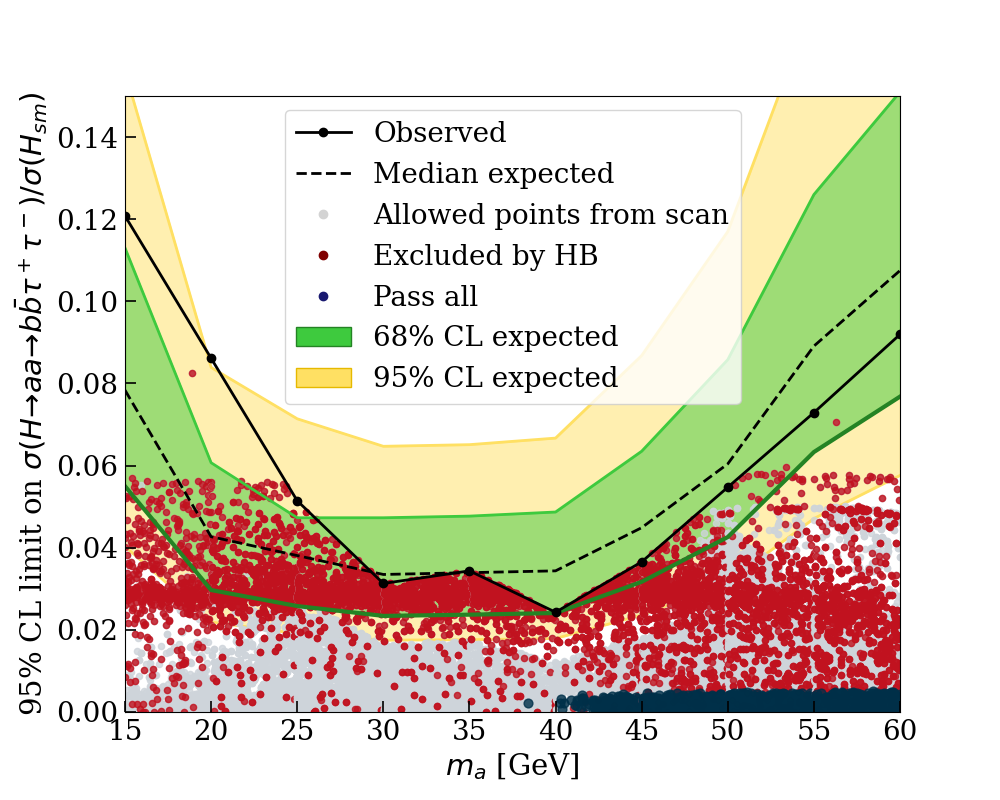}}
   \resizebox{0.4\textwidth}{0.21\textheight}{\includegraphics{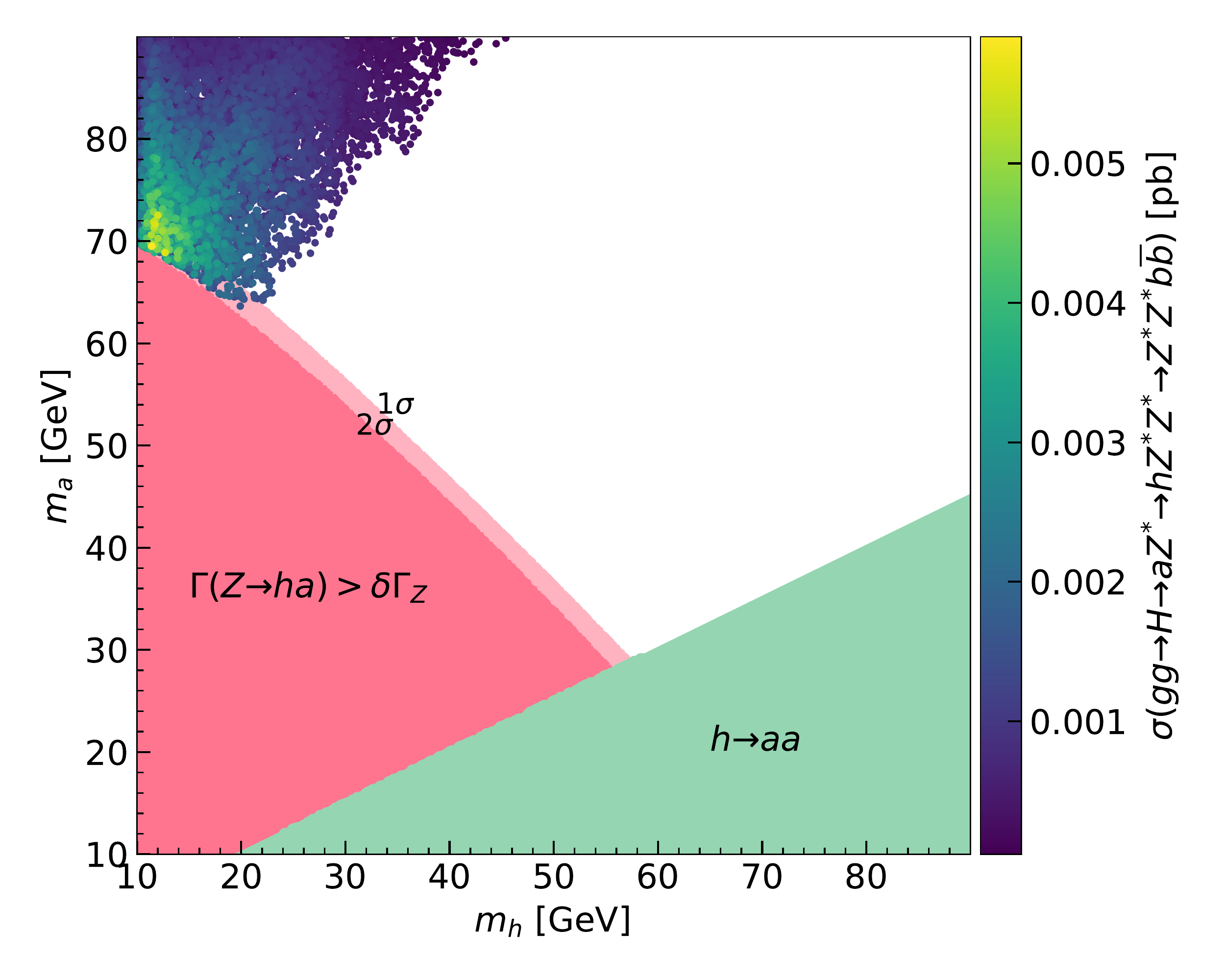}}
	\caption{(Left panel) observed and expected upper limits on $\sigma(H \rightarrow aa \rightarrow b\overline{b} \tau^+\tau^-)/\sigma_{\rm SM}(H)$~\cite{CMS:2018zvv} at 95\% C.L. (Right panel) $\sigma(gg \to H \to aZ^{*} \to h Z^{*}Z^{*} \to Z^{*}Z^{*}b\overline{b})$ at 95\% C.L in the 2HDM Type-I} 
	\label{fig2}
\end{figure}

Fig.~\ref{fig1} shows the combination of the scanned model parameters, which satisfies all theoretical and experimental constraints, whereas the white space corresponds to the case where any possible mass combination is forbidden by observed signatures in one or more existing experimental searches. The coloured dot indicates the channel search to which each point is sensitive to. It is visible that the mass combination in the top left corner is sensitive to LEP searches. As a result, an update of the LHC at Run 3 may not exclude it. We then investigate  the reinterpretation of $H \to aa \to b\overline{b}\tau^+\tau^-$ in the framework of the 2HDM Type-I,  see Fig.~\ref{fig2} (left panel). One can read that the model parameters with sensitivity to this search are already excluded by previous experimental searches (red points). An overall outcome, after testing $H \to aa(hh) \to b\overline{b}\mu^+\mu^-,\mu^+\mu^-\tau^+\tau^-$, is that viewing 2HDM Type-I as a reference framework for reinterpreting exotic Higgs decays searches in "traditional" modes is not advantageous. We, therefore, present in the right panel of Fig.~\ref{fig2} an alternative signature arising from $H \to aZ^{*} \to Z^{*}Z^{*}h$, with $Z^{*}$ being off-shell and $h\to bb$. The process yields a sizeable cross section\footnote{The LO cross section of Higgs production is computed by \texttt{Sushi}~\cite{Harlander:2012pb}} that could reach 0.06 pb. We will further look into the case where $Z^{*}Z^{*} \to \mu^+\mu^-jj$, leading to $\mu^+\mu^-jjb\overline{b}$ in the final state.

\section{Collider phenomenology}
\begin{figure*}[h!]
	\centering
	\resizebox{0.39\textwidth}{!}{
		\includegraphics{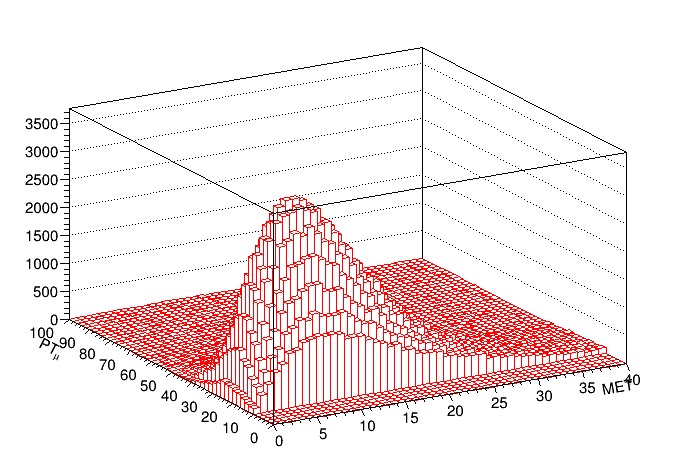}}
	\resizebox{0.39\textwidth}{!}{
		\includegraphics{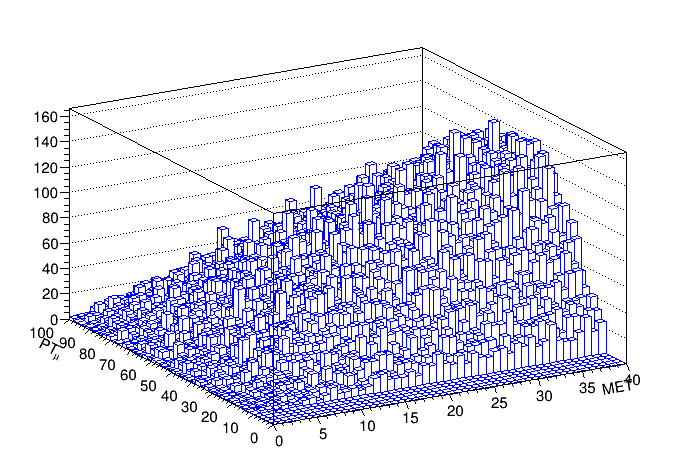}}
	\caption{Correlation between $p_T^\mu$ vs. $E_T$ for the signal (red) and background (blue) at the detector level}
	\label{fig3}
\end{figure*}

We use \texttt{MadGraph-v.9.2.5}~\cite{Alwall:2014hca} for event generation of signal and background\footnote{QCD corrections to both signal and background are taken trough K-factor~\cite{Bevilacqua:2011aa}}, \texttt{PYTHIA8}~\cite{Sjostrand:2006za} for showering and hadronisation, and \texttt{Delphes-3.5.0}~\cite{deFavereau:2013fsa} for detector simulation. The dominant background process arises from top pair production in association with two Initial State Radiation (ISR) jets, whereas the irreducible background, $pp \to Z^{*}Z^{*} bb \to \mu \mu jjbb$, is negligible. In this analysis, we rely on a di-muon trigger~\cite{CMS:2021yvr}, where the $p_T$ threshold of the leading(subleading) muon is 17(8) GeV, for events selection. We also require the following acceptance cuts at the detector level: $ p_T^{j,~b} > 20~\text{GeV},~p_T^l > 10~\text{GeV},~|\eta(l,b)|< 2.5,~|\eta(j)|< 5.0,~\Delta R > 0.4$. To favour the signal over the background, we examine different 2D distributions correlating the  missing transverse energy with various kinematic variables. As one can see from Fig.~\ref{fig3}, the signal and background distributions are anti-correlated and thus a cut on the missing transverse energy, $E_T <25$ GeV, will enhance the signal and suppress the background. Through similar reasoning, we require further cuts: $p_T^{j} < 75$ GeV, $p_T^{\mu} < 40$ GeV, $m_{\mu\mu} < 40~\text{GeV}$, $\Delta R (b_i, \mu_j)<2.5$, $\Delta R (\mu_1, \mu_2)<2.5$,  $\Delta R (j_1, j_2)<2.5$ and $\Delta R (b_1, b_2)<2.5$.

\begin{figure}[h!]
\centering
	\centering
\resizebox{0.35\textwidth}{!}{
	\includegraphics{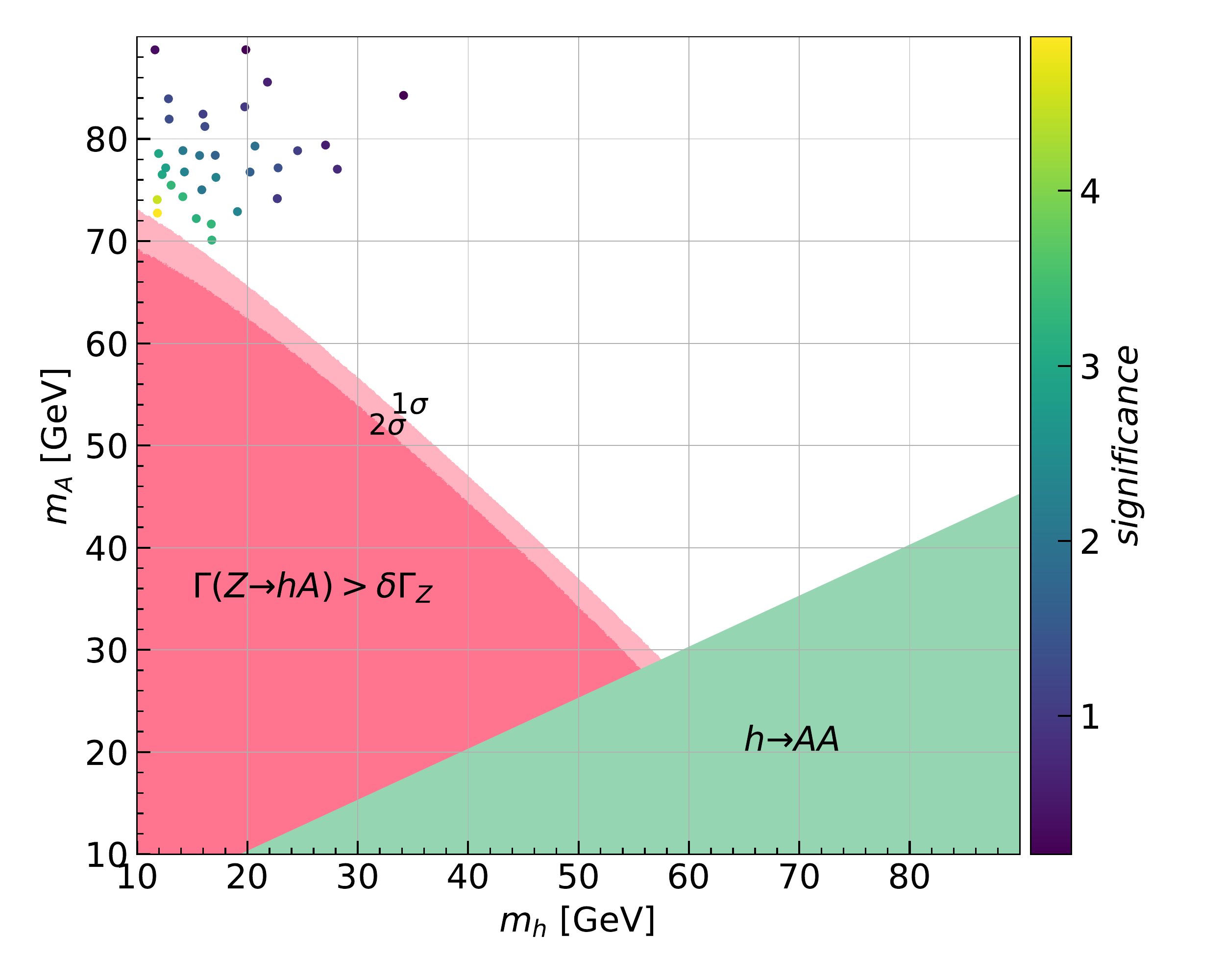}}
	\centering
\resizebox{0.35\textwidth}{!}{
	\includegraphics{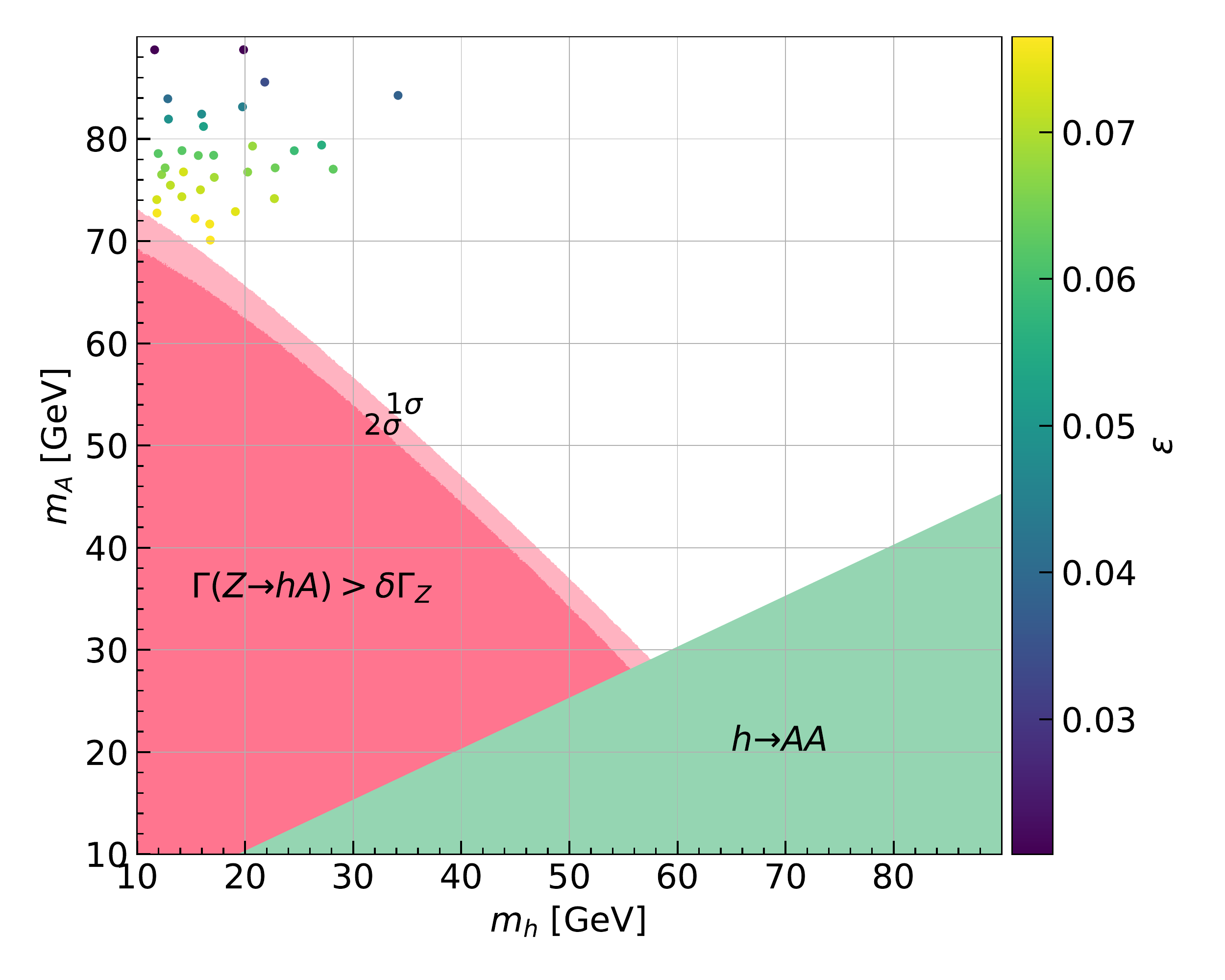}}
\caption{Significance ($\Sigma$) and efficiency ($\epsilon$) of each BP}
\label{fig4}
\end{figure}

We consider many Benchmark Points (BPs) to map the parameter space of the 2HDM Type-I. We show in
Fig.~\ref{fig4} the efficiency ($\epsilon$) and significance ($\Sigma$)\footnote{$\epsilon = \frac{\text{cross section after cuts}}{\text{cross section before cuts}}$ and $\Sigma=\frac{S}{\sqrt{B}}$, where $S$($B$) is the signal(background) yield after the discussed cutflow} of each BP, with while requiring the cut flow listed above. It
is interesting to note that there are points on the grid with a large significance for $\mathcal{L}=300~\text{fb}^{-1}$ (see Tab.~\ref{tab2}),  opening up the high possibility of accessing the above signature during Run-3 and providing clear evidence of discovery at the High-Luminosity LHC (HL-LHC).

\begin{table}[h!]
	\begin{center}
		\resizebox{0.7\textwidth}{!}{
			\begin{tabular}{||c|c|c|c|c|c|c|} \hline\hline
				BP & $m_h$ (GeV)  & $m_a$ (GeV) & $\sigma$ (pb) & $K$-factor  & significance ($\Sigma$)  & efficiency $(\epsilon)$ \\  
				\hline
				BP1 & 11.85 & 72.75 & $4.82\times 10^{-4}$ & 2.689    & \textcolor{black}{4.88} &  0.0758 \\
				\hline
				BP2 & 15.37 & 72.21 & $3.28\times 10^{-4}$ & 2.63   & \textcolor{black}{3.20}   & 0.0757 \\
				\hline
				BP3 & 17.15 & 76.24 & $2.54\times 10^{-4}$ &  2.63  & 2.29  & 0.0689 \\
				\hline
				BP4 & 13.09  & 75.47 & $3.538\times 10^{-4}$ & 2.65   & 3.31    &  0.0709 \\
				\hline
				BP5 & 14.15  & 74.35 & $3.458\times 10^{-4}$ & 2.62    &  \textcolor{black}{3.29}   & 0.072  \\
				\hline 
				BP6 & 11.96  & 78.57 & $3.557\times 10^{-4}$ & 2.69     & 2.97   &   0.062 \\
				\hline  
				BP7 & 14.16  & 78.86 & $2.572\times 10^{-4}$ &  2.648    &  2.11  & 0.062   \\
				\hline 
				BP8 & 11.83  & 74.06 & $4.577\times 10^{-4}$ & 2.69   & \textcolor{black}{4.51}  & 0.073  \\ 
				\hline \hline
			\end{tabular}
		}		
	\end{center}
	\caption{Significance ($\Sigma$) and efficiency ($\epsilon$) of some selected BPs for $\sqrt s=13$ TeV and ${\cal L}=300$ fb$^{-1}$~\cite{Moretti:2022fot}. Here, $\sigma$ indicates the signal LO cross section and  $K\text{-factor} (= \sigma^{\rm NNLO}/\sigma^{\rm LO} \sim 2.6-2.7)$ quantifies the NNLO QCD correction to the Higgs production. The NLO QCD correction to $ggt\overline{t} \sim -27\%$~\cite{Bevilacqua:2011aa}}
	\label{tab2}
\end{table}
\section{Conclusion}
Within the framework of the 2HDM Type-I, we have reviewed the search for exotic Higgs decays, $H \to aa(hh)$, in different final states, performed at Run 2 with an integrated luminosity of $35.9~\text{fb}^{-1}$ to draw the actual sensitivity of these experiments to the Type-I parameter space. In doing so, we have found that the accessible parameter space  is already excluded by previous experimental searches. We have thus suggested an alternative signature to search for light Higgses, $pp\to H \to Z^{*}a\to  Z^{*} Z^{*}h\to  \mu^+\mu^- jj b\overline{b}$, offering potential discovery at the LHC.

\section*{Acknowledgements}
The work of SM is supported in part through the NExT Institute and the STFC Consolidated Grant No. ST/L000296/1. SS is fully supported through the NExT Institute.

\end{document}